\documentclass[twocolumn,english,aps,prd,preprintnumbers,nofootinbib]{revtex4-1}
\usepackage{graphicx}
\usepackage{amsmath}
\usepackage{amssymb}
\usepackage{amsfonts}
\usepackage{hyperref}
\usepackage{url}
\usepackage{color}
\usepackage{bm}
\usepackage{array}

\newcommand{\be}{\begin{equation}}
\newcommand{\ee}{\end{equation}}
\newcommand{\ba}{\begin{eqnarray}}
\newcommand{\ea}{\end{eqnarray}}
\newcommand{\nn}{\nonumber}

\renewcommand{\[}{\begin{equation}}
\renewcommand{\]}{\end{equation}}

\def\lcdm{$\Lambda$CDM }
\makeatother

\begin{document}

\preprint{IFT-UAM/CSIC-19-130}

\title{What can machine learning tell us about the background expansion of the Universe?}

\author{Rub\'{e}n Arjona}
\email{ruben.arjona@uam.es}

\author{Savvas Nesseris}
\email{savvas.nesseris@csic.es}

\affiliation{Instituto de F\'isica Te\'orica UAM-CSIC, Universidad Auton\'oma de Madrid,
Cantoblanco, 28049 Madrid, Spain}

\date{\today}

\begin{abstract}
Machine learning (ML) algorithms have revolutionized the way we interpret data in astronomy, particle physics, biology and even economics, since they can remove biases due to a priori chosen models. Here we apply a particular ML method, the genetic algorithms (GA), to cosmological data that describes the background expansion of the Universe, namely the Pantheon Type Ia supernovae and the Hubble expansion history $H(z)$ datasets. We obtain model independent and nonparametric reconstructions of the luminosity distance $d_L(z)$ and Hubble parameter $H(z)$ without assuming any dark energy model or a flat Universe. We then estimate the deceleration parameter $q(z)$, a measure of the acceleration of the Universe, and we make a $\sim4.5\sigma$ model independent detection of the accelerated expansion, but we also place constraints on the transition redshift of the acceleration phase $(z_{\textrm{tr}}=0.662\pm0.027)$. We also find a deviation from \lcdm at high redshifts, albeit within the errors, hinting toward the recently alleged tension between the SnIa/quasar data and the cosmological constant \lcdm model at high redshifts $(z\gtrsim1.5)$. Finally, we show the GA can be used in complementary null tests of the \lcdm via reconstructions of the Hubble parameter and the luminosity distance.
\end{abstract}
\maketitle

\section{Introduction}
Cosmology has reached a stage of near percent level precision with a wide range of theoretical models that describe rigorous and accurate measurements. However, the explanation as to why the Universe is undergoing a period of accelerated expansion still remains an open question and the cause of this phenomenon is usually attributed to a dark energy (DE) component \cite{Copeland:2006wr}. The standard cosmological model contains the cosmological constant $\Lambda$ and a cold dark matter component ($\Lambda$CDM) \cite{Peebles:2002gy} and is at the moment the best candidate to explain the accelerated expansion of the Universe as it is in excellent agreement with all of the current data \cite{Aghanim:2018eyx}.

However, there is a plethora of other models as well, many of which are included in the pipelines of upcoming surveys, such as Euclid \cite{Amendola:2016saw}. These models range from canonical scalar fields \cite{Ratra:1987rm,Wetterich:1987fm,Caldwell:1997ii}, scalar fields with a generalized kinetic terms \cite{ArmendarizPicon:2000dh,ArmendarizPicon:2000ah} or a nonminimal couplings \cite{Uzan:1999ch,Perrotta:1999am,Riazuelo:2001mg} in addition to general relativity (GR), coupled DE models \cite{Dent:2008vd}, modifications of the Einstein-Hilbert action \cite{Starobinsky:1980te}, the Chaplygin gas \cite{Bento:2002ps} or extra dimensions \cite{Deffayet:2001pu}. For further reviews see \cite{Amendola:2015ksp,Tsujikawa:2010zza,Nojiri:2006ri,Sotiriou:2008rp,Nojiri:2010wj,DeFelice:2010aj,Amendola:2016saw,Nojiri:2017ncd}.

This huge landscape of DE models makes the interpretation of the cosmological observations difficult as the results, e.g. the value of the matter content of the Universe $\Omega_{m0}$, depend on the particular model chosen. For example, the Planck mission provides an accurate value for the matter density parameter today $\Omega_{m0}=0.315\pm0.007$, see Ref.~\cite{Aghanim:2018eyx}, however this value is specific to the \lcdm model as it was obtained assuming the \lcdm model to be the correct theory, hence is model dependent. To remove biases due to choosing an \emph{a priori} defined model, it is important to use reconstruction techniques and model independent approaches, see for example \cite{Nesseris:2010ep}. One such approach is the use of machine learning (ML) methods, which has already lead to many successes in cosmology \cite{Ntampaka:2019udw}. ML methods have been used to reduce the scatter in cluster mass estimates \cite{Ho:2019zap}, to distinguish between standard and modified gravity theories from statistically similar weak lensing maps \cite{Peel:2018aei}, and have been found to be useful for the next generation CMB experiments \cite{Caldeira:2018ojb}, N-body simulations \cite{He:2018ggn}, cosmological parameters inference \cite{Ravanbakhsh:2016xpe}, dark energy model comparison \cite{Escamilla-Rivera:2019hqt}, supernova classification \cite{Narayan:2018rxv} and strong lensing probes \cite{Lanusse:2017vha}.

In this paper we will apply a particular ML method, the genetic algorithms (GA), which can be described as a stochastic search approach. The GA have been used in many disciplines ranging from astrophysics, e.g. to determine the photometric redshift \cite{Hogan:2014qsa}, to find the optimum parameters for cosmic ray injection and propagation \cite{Luo:2019qbk}, to fit dusty galaxies \cite{DeGeyter:2012tf},
to perform galaxy classification \cite{Calleja:2004xxx}, in particle physics to constrain the MSSM \cite{Akrami:2009hp,Allanach:2004my} or resonances in Lambda reactions \cite{Ireland:2003yg}, but also in finance \cite{Geneticfinance:2002xxx,Geneticfinance:2014xxx} and biology \cite{Dikhil2016}. More recently, they have also been applied to cosmology for data reconstruction \cite{Arjona:2020doi,Arjona:2020kco,Nesseris:2012tt,Nesseris:2014qca,Nesseris:2013bia,Bogdanos:2010yac,Nesseris:2010ep,Nesseris:2010zp,Bogdanos:2009ib}. One of the most effective use of these methods is the reconstruction of null tests, i.e. pass/fail test made of variables of a theory that should always be constant for all values of the parameters, and can be used to test theories in a model independent way.

In light of the near future experiments that will gather a vast amount of data, such as Euclid and LSST, it is necessary to perform model independent tests to check for possible tensions that could be due to systematics or new physics. Specifically, null tests for \lcdm have already been applied to the cosmological constant model  \cite{Sahni:2008xx,Zunckel:2008ti,Nesseris:2010ep}, interacting DE models \cite{vonMarttens:2018bvz}, the growth-rate data \cite{Nesseris:2014mfa}, the cosmic curvature \cite{Yahya:2013xma,Cai:2015pia} and to probe the scale-independence of the growth of structure in the linear regime \cite{Franco:2019wbj}.

The structure of our paper is a follows. In Sec.~\ref{sec:ga} we present the theoretical background of the GA approach we use in our analysis, in Sec.~\ref{sec:analysis} we present our reconstructions and the results on the deceleration parameter, the transition redshift and the two null tests based on the Hubble parameter and luminosity distance. Finally, in Sec.~\ref{sec:conclusions} summarize our results and present our conclusions.

\section{The Genetic Algorithms \label{sec:ga}}
Here we present the theoretical background of the implementation of the GA in our analysis.

Machine learning is a subset of artificial intelligence that aims to build mathematica models that describe a given set of data. One particular class of ML methods are the genetic algorithms (GA) which specialize in unsupervised symbolic regression of data. This means that the GA can reconstruct an analytic function that describes the data, using one or more variables. The GA achieve this by mimicking from biology the notion of evolution. In particular, this is expressed via natural selection and the genetic operations of crossover and mutation. Hence, a group of candidate functions evolves over time under under pressure from the stochastic operators of crossover, i.e. the merging of two or more individuals that form an offspring, and mutation, i.e. a random change in the genetic makeup of an individual.

The ``reproductive" success of a population, or in other words its probability that it will produce offspring, is usually taken to be proportional to its fitness, which expresses how well every individual agrees with the data. In our analysis, this is quantified via a usual $\chi^2$ statistic, which we discuss in detail in the next section for the data we will use. For more details on the GA and various applications to cosmology see \cite{Bogdanos:2009ib,Nesseris:2012tt,Arjona:2020kco}.

In this analysis we consider the Pantheon Type Ia Supernovae (SnIa) and $H(z)$ data sets, so in practice, the procedure to reconstruct them, proceeds as follows. First, we choose an orthogonal basis of functions, traditionally called the ``grammar", with which an initial population of functions is constructed. These function are randomly picked so that every member of the population codifies an initial guess for both the luminosity distance $d_\textrm{L}(z)$ and the Hubble parameter $H(z)$. While this choice for the grammar might seem crucial for the symbolic regression, it has been shown in Ref.~\cite{Bogdanos:2009ib} that it only affects the rate of convergence of the GA.

In this first step one may also impose any necessary physical priors, for example that the value of the Hubble parameter today is $H(z=0)=H_0$ or that the luminosity distance at $z=0$ is zero, i.e. $d_\textrm{L}(z=0)=0$. This step is important as we want to avoid any unphysical functions that could unnecessarily delay the convergence of our GA code. We also demand that all functions reconstructed by the GA are continuous and differentiable, without any singularities in the redshift range covered by the data, so as to avoid overfitting or any spurious reconstructions. These are the only physical assumptions we do and we make no assumption on any particular DE or modified gravity model or even on the curvature of the Universe.

After the initial population has been constructed, the fitness of each member is estimated by a $\chi^2$ statistic, using as input the SnIa and $H(z)$ data. Afterwards, using a tournament selection, see Ref.~\cite{Bogdanos:2009ib} for more details, the best-fitting functions in every generation are chosen and the two stochastic operations of the crossover and the mutation are applied. In order to ensure convergence this procedure is then repeated hundreds of times and with various random seeds, so as to properly explore the functional space.

The final output of the GA code is a couple of two continuous and differentiable functions of the redshift $z$ that describe the Hubble parameter $H(z)$ and the luminosity distance $d_\textrm{L}(z)$ respectively. However, the GA on its own does not provide any estimate of the errors of the reconstructed functions, something which is necessary for the statistical interpretation of the data. To do so, we implement the path integral approach of Refs.~\cite{Nesseris:2012tt,Nesseris:2013bia}, where the errors are estimated by calculating analytically a path integral over all functions that may be surveyed by the GA. This error reconstruction method has been exhaustively examined and compared against a bootstrap Monte-Carlo by Ref.~\cite{Nesseris:2012tt}.

At this point it should be noted that while no assumptions on a particular cosmological model, such as the $\Lambda$CDM were made, sometimes the data themselves may not be completely model-independent. One such example is the JLA SnIa compilation \cite{Betoule:2014frx}, for which one has to fit the cosmological parameters at the same time with the light-curve parameters, which are of astrophysical origin. Another similar case is that of the Pantheon compilation \cite{Scolnic:2017caz}, for which some model dependence may still remain, despite that the light-curve parameters have already been integrated over. The reason for this is that the SnIa surveys have to take into account specific corrections regarding the peculiar velocities, assuming linear theory and the \lcdm model \cite{Mohayaee:2020wxf}. Moreover, a fiducial background model is typically assumed in order to derive the covariance matrix of the data \cite{Scolnic:2017caz}. In our case we can safely assume that these effects have a very small effect on the reconstruction process as the best-fit is close to the \lcdm model.

Finally, other non-parametric approaches to data reconstruction include the Gaussian processes (GP), which are based on the assumption that the data is described by a stochastic Gaussian process that can be mapped to a cosmological function of interest. For recent applications of GP to cosmology see \cite{Shafieloo:2012ht,Busti:2014aoa,Pinho:2018unz,Bengaly:2019oxx}. The GP require the choice of a kernel function and that of a fiducial model, usually taken to be $\Lambda$CDM, although in Ref.~\cite{Shafieloo:2012ht} it is claimed that both of these choices do not influence the reconstruction. On the other hand, the GA require no prior assumptions, eg of a DE model or a flat Universe, besides the choice of the grammar which only affects the rate of convergence \cite{Bogdanos:2009ib}. Qualitatively, by comparing plots of the same reconstructed parameter we find that both the GA and GP give similar errors, e.g. see the reconstruction of $H(z)$ in Fig 2 of Ref.~\cite{Pinho:2018unz}.

\section{Analysis and results \label{sec:analysis}}
\subsection{The data \label{sec:data}}
The null tests we will consider here are the $\textrm{Om}(z)$ statistic \cite{Sahni:2008xx,Zunckel:2008ti} and a new null test derived from the luminosity distance, that we present here for the first time. We thus propose applying ML methods, in particular the GA, to fit to the Pantheon Type Ia supernovae (SnIa) data compilation \cite{Scolnic:2017caz} and the $H(z)$ data compilation of Ref.~\cite{Arjona:2018jhh} to obtain a model independent reconstruction of the luminosity distance $d_L(z)$ and of the Hubble parameter $H(t)\equiv \frac{\dot{a}}{a}$, where $a(t)$ is the scale factor in the Robertson-Walker metric, and the dot stands for a derivative with respect to the cosmic time $t$.

\begin{table}[!t]
\caption{The $H(z)$ data used in our analysis (in units of $\textrm{km}~\textrm{s}^{-1} \textrm{Mpc}^{-1}$). This compilation is partly based on those of Refs.~\cite{Moresco:2016mzx} and \cite{Guo:2015gpa}.\label{tab:Hzdata}}
\small{
\centering
\begin{tabular}{cccccccccc}
\\
\hline\hline
$z$  & $H(z)$ & $\sigma_{H}$ & Ref.   \\
\hline
$0.07$    & $69.0$   & $19.6$  & \cite{Zhang:2012mp}  \\
$0.09$    & $69.0$   & $12.0$  & \cite{STERN:2009EP} \\
$0.12$    & $68.6$   & $26.2$  & \cite{Zhang:2012mp}  \\
$0.17$    & $83.0$   & $8.0$   & \cite{STERN:2009EP}    \\
$0.179$   & $75.0$   & $4.0$   & \cite{MORESCO:2012JH}   \\
$0.199$   & $75.0$   & $5.0$   & \cite{MORESCO:2012JH}   \\
$0.2$     & $72.9$   & $29.6$  & \cite{Zhang:2012mp}   \\
$0.27$    & $77.0$   & $14.0$  & \cite{STERN:2009EP}   \\
$0.28$    & $88.8$   & $36.6$  & \cite{Zhang:2012mp}  \\
$0.35$    & $82.7$   & $8.4$   & \cite{Chuang:2012qt}   \\
$0.352$   & $83.0$   & $14.0$  & \cite{MORESCO:2012JH}   \\
$0.3802$  & $83.0$   & $13.5$  & \cite{Moresco:2016mzx}   \\
$0.4$     & $95.0$   & $17.0$  & \cite{STERN:2009EP}    \\
$0.4004$  & $77.0$   & $10.2$  & \cite{Moresco:2016mzx}   \\
$0.4247$  & $87.1$   & $11.2$  & \cite{Moresco:2016mzx}   \\
$0.44$    & $82.6$   & $7.8$   & \cite{Blake:2012pj}   \\
$0.44497$ & $92.8$   & $12.9$  & \cite{Moresco:2016mzx}   \\
$0.4783$  & $80.9$   & $9.0$   & \cite{Moresco:2016mzx}   \\
\hline\hline
\end{tabular}
\begin{tabular}{cccccccccc}
\\
\hline\hline
$z$  & $H(z)$ & $\sigma_{H}$ & Ref.   \\
\hline
$0.48$    & $97.0$   & $62.0$  & \cite{STERN:2009EP}   \\
$0.57$    & $96.8$   & $3.4$   & \cite{Anderson:2013zyy}   \\
$0.593$   & $104.0$  & $13.0$  & \cite{MORESCO:2012JH}  \\
$0.60$    & $87.9$   & $6.1$   & \cite{Blake:2012pj}   \\
$0.68$    & $92.0$   & $8.0$   & \cite{MORESCO:2012JH}    \\
$0.73$    & $97.3$   & $7.0$   & \cite{Blake:2012pj}   \\
$0.781$   & $105.0$  & $12.0$  & \cite{MORESCO:2012JH} \\
$0.875$   & $125.0$  & $17.0$  & \cite{MORESCO:2012JH} \\
$0.88$    & $90.0$   & $40.0$  & \cite{STERN:2009EP}   \\
$0.9$     & $117.0$  & $23.0$  & \cite{STERN:2009EP}   \\
$1.037$   & $154.0$  & $20.0$  & \cite{MORESCO:2012JH} \\
$1.3$     & $168.0$  & $17.0$  & \cite{STERN:2009EP}   \\
$1.363$   & $160.0$  & $33.6$  & \cite{Moresco:2015cya}  \\
$1.43$    & $177.0$  & $18.0$  & \cite{STERN:2009EP}   \\
$1.53$    & $140.0$  & $14.0$  & \cite{STERN:2009EP}  \\
$1.75$    & $202.0$  & $40.0$  & \cite{STERN:2009EP}  \\
$1.965$   & $186.5$  & $50.4$  & \cite{Moresco:2015cya}  \\
$2.34$    & $222.0$  & $7.0$   & \cite{Delubac:2014aqe}   \\
\hline\hline
\end{tabular}}
\end{table}

In our analysis we use $1048$ data points in the range $z\in [0,2.26]$, along with their covariances, from the Pantheon set \cite{Scolnic:2017caz}, and $36$ points in the range $z\in[0,2.34]$ from the $H(z)$ compilation, presented in Table~\ref{tab:Hzdata}. On the other hand, we make no assumptions for $H_0$ and derive it directly from the $H(z)$ data, as we will see later on. Measurements of the Hubble expansion $H(z)$ data are performed either by the differential age method or by the clustering of galaxies or quasars. The former is possible due to the redshift drift of distant objects over a decade or longer, since in GR the $H(z)$ can also be expressed via the rate of change of the redshift $H(z)=-\frac{1}{1+z}\frac{dz}{dt}$ \cite{Jimenez:2001gg}. The latter is related to the clustering of galaxies or quasars and it leads to measurements of $H(z)$ by measuring the radial BAO peak \cite{Gaztanaga:2008xz}.

The $H(z)$ data measured via the differential age method are obtained by following the differential evolution of $D_n$4000, a spectral feature of very massive and passive galaxies. The main source of systematics is the astrophysical modelling of the stellar metallicity, namely via the M11 and BC03 models discussed in Ref.~\cite{Moresco:2016mzx}. However, by implementing strict selection criteria it was shown in Ref.~\cite{Moresco:2016mzx} that it is possible to keep the systematics under control. Furthermore, the $H(z)$ data are independent of any cosmology-based constraint, i.e. a fiducial cosmological model, they are assumed to be uncorrelated with each other and share no correlations with the SnIa data \cite{Moresco:2016mzx}.

Finally, for the likelihood for the $H(z)$ data we use a standard $\chi^2$, given by
\be
\chi^2_\textrm{H}=\sum_{i=1}^N \left(\frac{H_i-H_{\textrm{GA}}(z_i)}{\sigma_i}\right)^2,
\ee
while for the SnIa data we use the expressions found in Appendix C of Ref.~\cite{Conley:2011ku}. Note that in our analysis we use the two data sets in the following manner: we use the $H(z)$ data for the deceleration parameter $q(z)$ and the $\textrm{Om}_\textrm{H}$ null test, while for the $\textrm{Om}_\textrm{dL}$ null test we will use the SnIa data.

\subsection{The reconstructions \label{sec:recon}}
We reconstruct the Hubble parameter by applying the GA to the $H(z)$ data, while the value of the Hubble parameter $H_0$ was derived through minimizing the $\chi^2$ analytically as the $\chi^2$ is quadratic in $H_0$, see Ref.~\cite{Basilakos:2018arq}. For the SnIa, due to the degeneracy between the absolute magnitude $M$ and the Hubble parameter $H_0$, we used the value extracted from the $H(z)$ data, given below. In both cases, no assumptions such as a flat Universe or a specific DE model were made, hence our results are almost completely model independent.

Note that sometimes the data are themselves model dependent, with an infamous example being the SnIa, as one must optimize parameters in the lightcurve function simultaneously with those of the assumed model. Furthermore, a covariance matrix is typically inferred based on an assumed background model, usually $\Lambda$CDM. However, since in our case the best-fit is close to \lcdm and the errors are much larger than the effects of the model-bias in the covariance, we can safely assume for now that these effects have a minimal effect to the minimization.

\begin{figure}[!t]
\centering
\includegraphics[width = 0.48\textwidth]{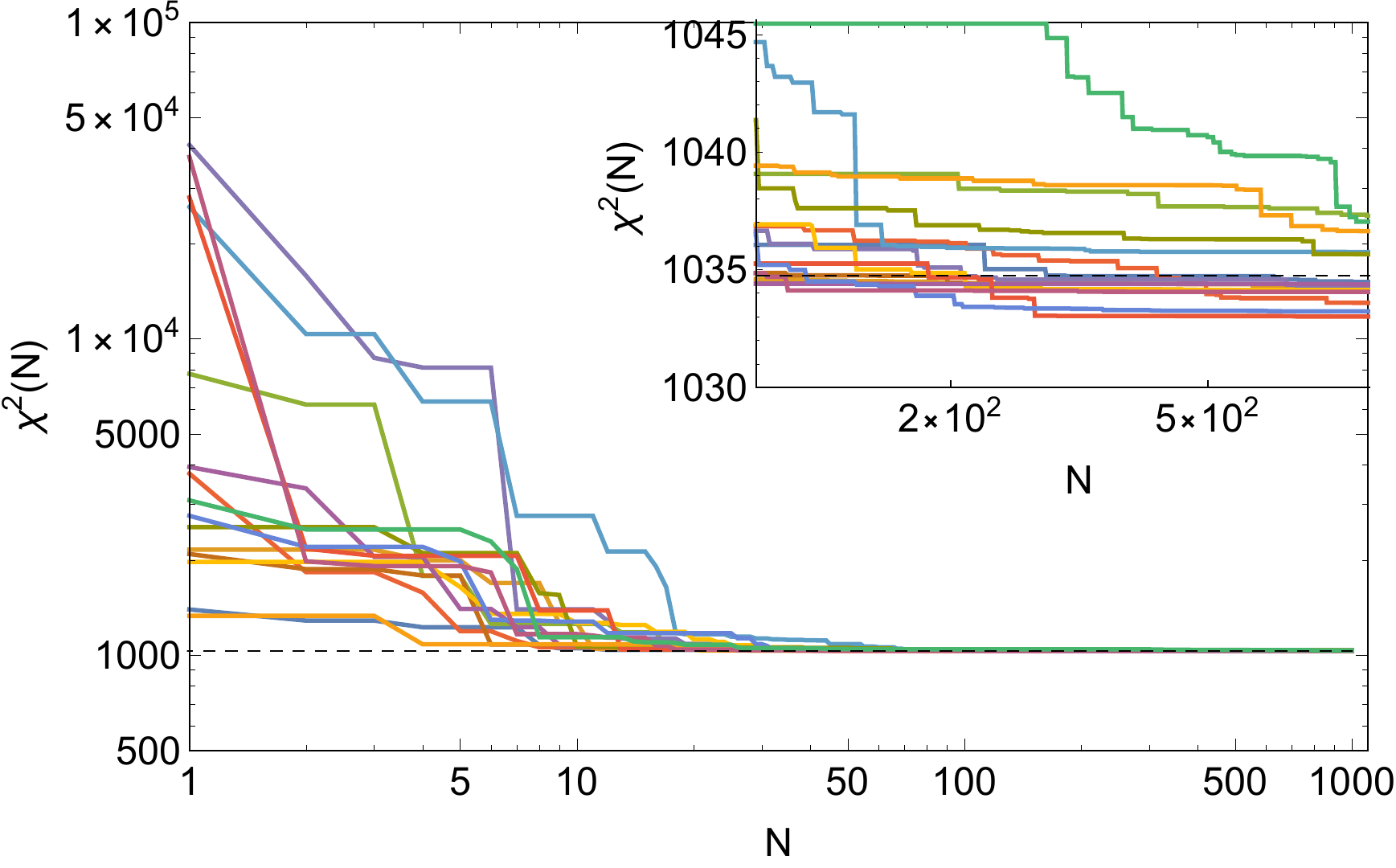}
\caption{The genetic evolution of several different initializations of the GA code with different seed random numbers for the SnIa data as a function of the generation number. In most cases the GA has converged very quickly in the evolutionary history and reaches a lower $\chi^2$ than \lcdm does. \label{fig:snchi2}}
\end{figure}

In order to make sure we are not biasing our analysis due to the specific value of the random seed we have performed several simulations with different random seed numbers. We have also demanded that all functions, along with their derivatives, are continuous and have no singularities in the range covered by the data. As an example, the genetic evolution of several different initializations of the GA code with different seed random numbers for the SnIa data as a function of the generation number can be seen in Fig.~\ref{fig:snchi2}. In most cases, the GA has converged very quickly in the evolutionary history and in the majority of cases, the obtained $\chi^2$ is smaller than that of the \lcdm model.

\begin{figure*}[!t]
\centering
\includegraphics[width = 0.49\textwidth]{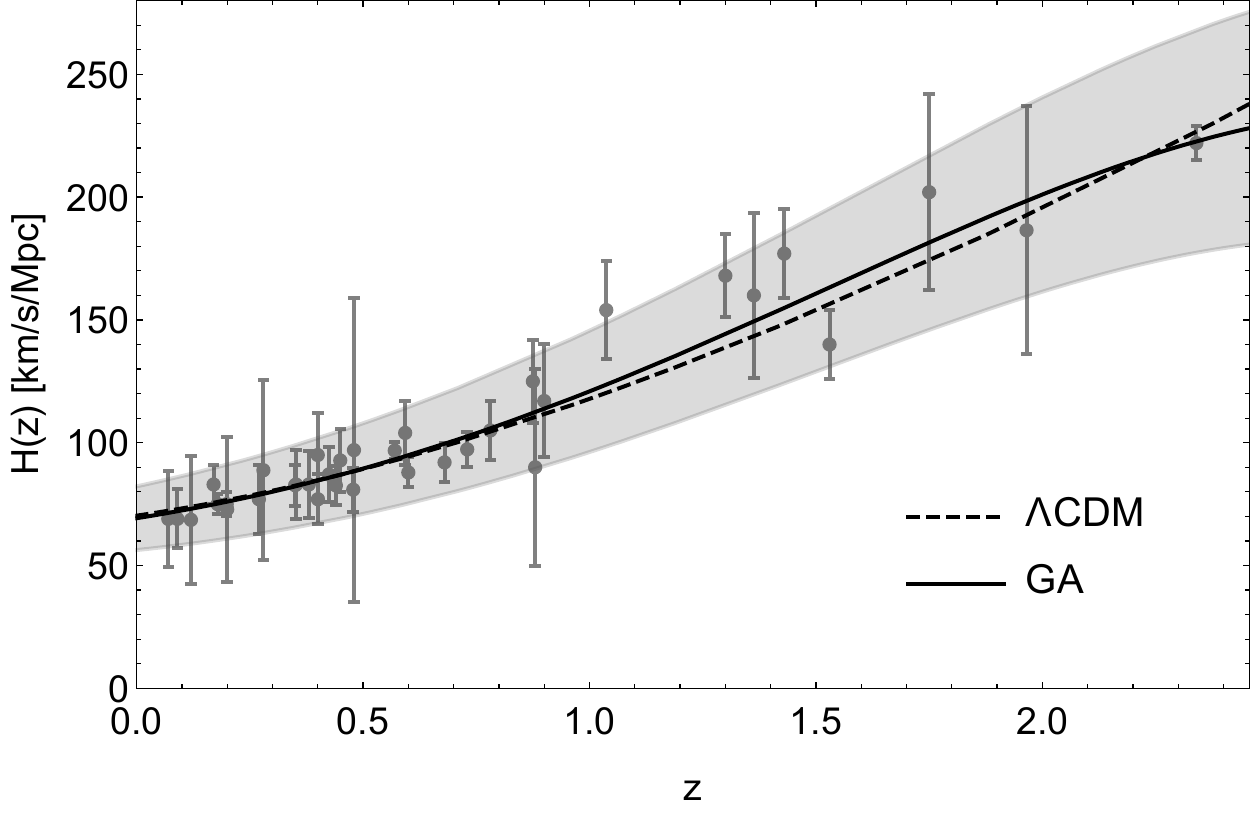}
\includegraphics[width = 0.49\textwidth]{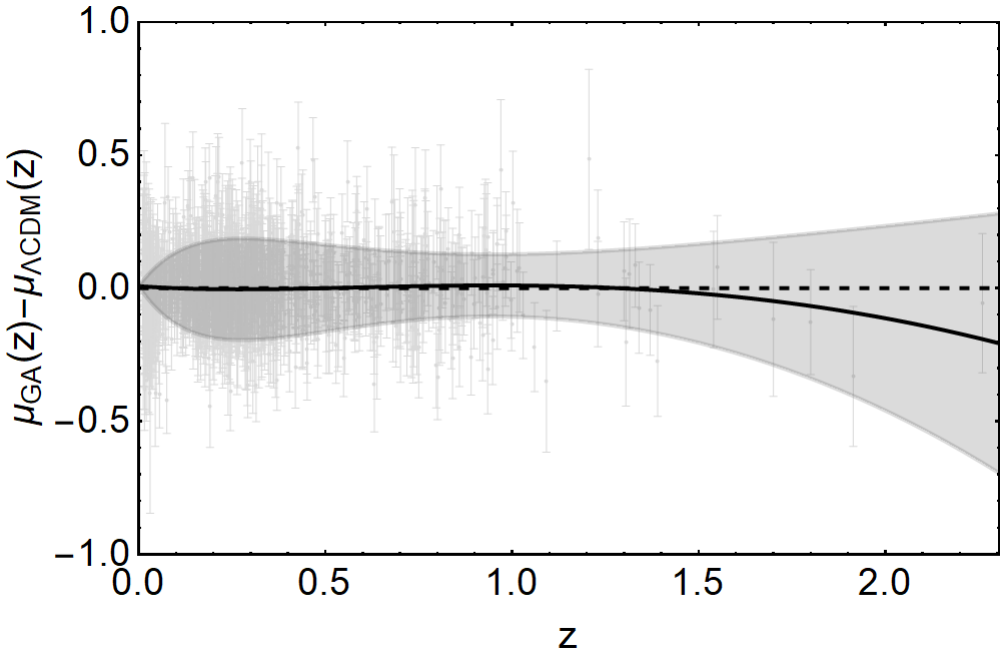}
\caption{Left: The $H(z)$ data compilation along with the \lcdm best-fit (dashed line) and the GA best-fit (solid black line). Right: The difference between the GA best-ft distance modulus of the Pantheon SnIa data (black line) and that of the \lcdm model (dashed line). The Pantheon SnIa data are shown as grey points in the background. \label{fig:Data}}
\end{figure*}

Following this approach and taking into account the constraints mentioned earlier, we find the best-fit GA functions to be
\ba
H_{0}&=&(69.27\pm12.00)~\textrm{km/s/Mpc},\label{eq:H0GA}\\
H(z)&=&H_{0}\left(1+z \left(0.652+0.228 z-0.017 z^3\right)^2\right),\label{eq:HzGA}~~~~\\
d_{L}(z)&=&\frac{c}{H_{0}}z\left(1+z\left(-0.054z-0.146e^{0.347z}+0.999\right)^2\right),\label{eq:dLGA} \nn\\
\ea
where $c$ is the speed of light and the constraint on $H_0$ was derived directly from the $H(z)$ data. The best-fit $\chi^2$ for the GA and \lcdm models are given in Table~\ref{tab:chi2}, while plots of the Hubble parameter and the distance modulus $\mu(z)=5 \log_{10}\left(d_L(z)/\textrm{Mpc}\right)+25$ versus \lcdm and the data are given in Fig.~\ref{fig:Data}. The agreement with the best-fit \lcdm model $(\Omega_{m0}=0.299\pm0.022)$ for the distance modulus $\mu(z)$ is at a subpercent level with \lcdm until $z\sim1.5$, but then it deviates similarly, albeit within the errors, to the reconstruction of Refs.~\cite{Risaliti:2018reu,Lusso:2019akb} that used SnIa and quasar data.

In order to make sure that the observed deviation from \lcdm is not affected by the choice of the particular dataset, we removed the last two points at high redshifts (at $z=1.914$ and $z=2.26$) of the Pantheon SnIa compilation in order to test the robustness of our results. We found that the GA fit is actually unaffected, with the $\chi^2$ values being respectively $\chi^2=1033.2$ for the \lcdm and $\chi^2=1032.94$ for the GA best-fit, where the latter in this case was found to be
\be
d_{L}(z)=\frac{c}{H_{0}}z\left(1+z(0.871-0.131z-0.001z^4)^2\right).
\ee
Specifically, we find that for the original dataset the difference of the distance moduli at $z=2.305$ is  $\mu_{\textrm{GA}} - \mu_{\Lambda \textrm{CDM}} = -0.200284$, while after removing the last two points we have $\mu_{\textrm{GA}} - \mu_{\Lambda \textrm{CDM}} = -0.246619$. In order to verify that our analysis is indeed robust, we extended it by repeated removing two random points, at any redshift this time, and then ran the reduced data set over the GA pipeline. In all cases we found that the behavior of the GA best-fit remains unchanged.

Therefore, since the residuals in the reduced dataset are clearly consistent with the ones of the complete set, as shown in the right panel of Fig.~\ref{fig:Data}, we are confident our analysis is robust and is not affected by the choice of the specific dataset. Thus, having determined the functional forms of $H(z)$ and the luminosity distance, we can now use them to place model independent tests on the background expansion of the Universe and reconstruct null tests of the \lcdm model.

\begin{table}[!t]
\caption{The $\chi^2$ for \lcdm and GA using the Pantheon SnIa and $H(z)$ data.\label{tab:chi2}}
\begin{centering}
\begin{tabular}{ccc}
 & SnIa & $H(z)$  \\\hline
$\chi^2_{\Lambda \text{CDM}}$  & 1034.73 & 19.476\\\hline
$\chi^2_{GA}$  & 1034.30 &  17.683\\\hline
\end{tabular}
\par
\end{centering}
\end{table}

The most critical parameter in determining whether the Universe is accelerating or not, is the  deceleration parameter which is given by
\ba
q(z)&=&-\frac{\ddot{a} a}{\dot{a}^2}\nn\\
&=&-1+(1+z)\frac{H'(z)}{H(z)},\label{eq:qz1}
\ea
where dots stand for derivatives with respect to the cosmic time $t$, while primes for derivatives with respect to the redshift $z$, where $a(t)=\frac{1}{1+z}$. The advantage of this parameter over the DE equation of state $w(z)$ is that the former only requires the knowledge of $H(z)$ and not that of cosmological parameters such as $\Omega_{m0}$ \cite{Nesseris:2010ep}.

\begin{figure}[!t]
\centering
\includegraphics[width = 0.49\textwidth]{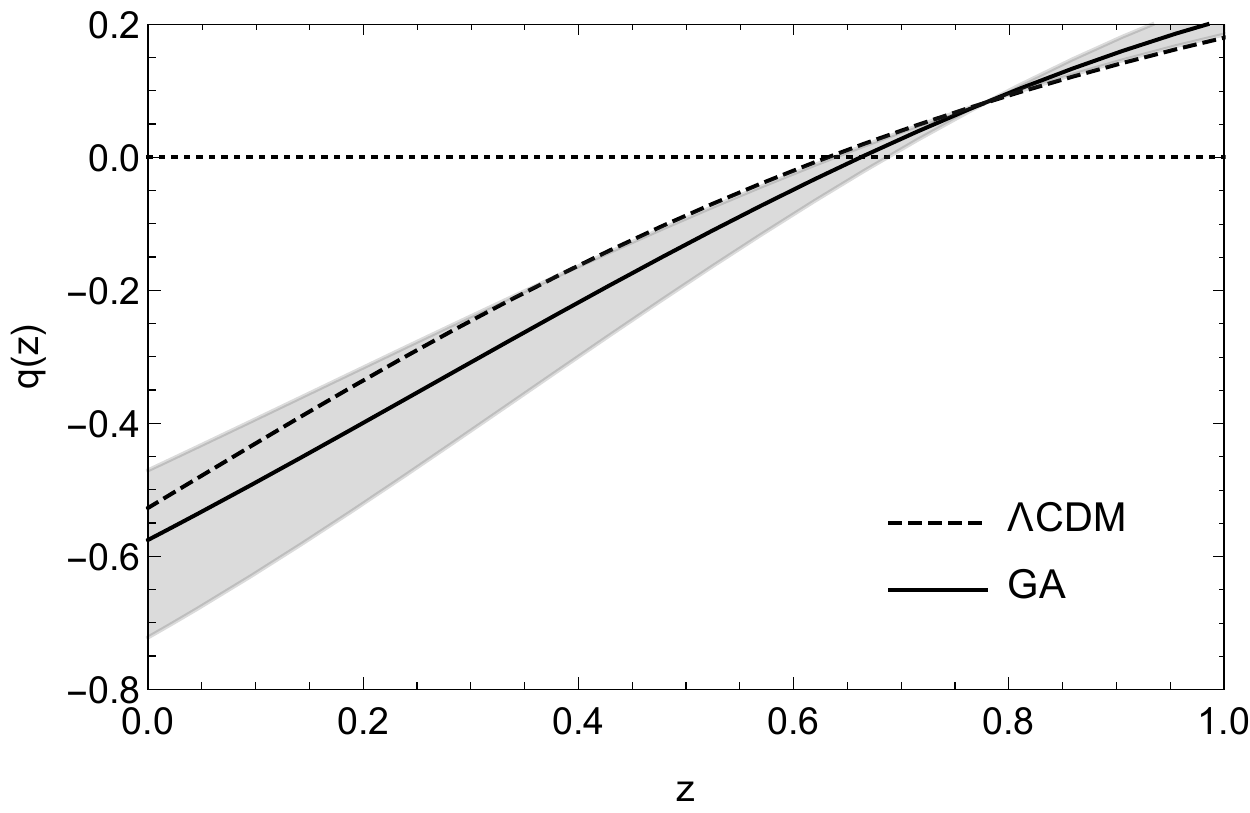}
\caption{The deceleration parameter given by Eq.~\eqref{eq:qz1} as reconstructed by using Eq.~\eqref{eq:HzGA}. The shaded region corresponds to the $1\sigma$ errors, while the transition redshift $z_{\textrm{tr}}$ corresponds to the point where $q(z)$ crosses zero. \label{fig:Hzqz}}
\end{figure}

For the Universe to accelerate today, we require (due to historical reasons) that $q_0<0$, e.g. for the \lcdm model we have $q_{0,\Lambda \textrm{CDM}}=-1+3\Omega_{m0}/2\simeq -0.528\pm0.011$ for the Planck best-fit $\Omega_{m0}=0.315$ \cite{Aghanim:2018eyx} and $q_{0,\Lambda \textrm{CDM}}=-0.613\pm 0.043$ for the \lcdm best-fit to the $H(z)$ data of $\Omega_{m0}=0.258\pm0.029$. Using the GA reconstruction of the Hubble parameter given by Eq.~\eqref{eq:HzGA} we can calculate the deceleration parameter given by Eq.~\eqref{eq:qz1} and the result is given in Fig.~\ref{fig:Hzqz}. The present value of the deceleration parameter is found to be $q_0\equiv q(z=0)=-0.575\pm0.132$, a $\sim4.5\sigma$ detection of the accelerated expansion of the Universe in a model-independent way.

We can also estimate the value of the transition redshift, i.e. the redshift where the deceleration parameter changes sign, see Refs~\cite{Capozziello:2014dwa,Farooq:2016zwm,Rani:2015lia,Yu:2017iju,Jesus:2017cyo,Jesus:2019nnk} for a list of recent estimates. From the GA reconstruction we find that $z_{\textrm{tr}}=0.662\pm 0.027$, while for the \lcdm the latter is equal to $z_{\textrm{tr},\Lambda \textrm{CDM}}=-1+2^{1/3}\left(\Omega_{m0}^{-1}-1\right)^{1/3}= 0.632\pm0.018$ for Planck and $z_{\textrm{tr},\Lambda \textrm{CDM}}=0.791\pm0.091$ for the $H(z)$ \lcdm best-fit. While the precision of these measurements seems worse than that of $\Lambda$CDM, in our case we have made very minimal assumptions and have not assumed any DE model.

\begin{figure*}[!t]
\centering
\includegraphics[width = 0.49\textwidth]{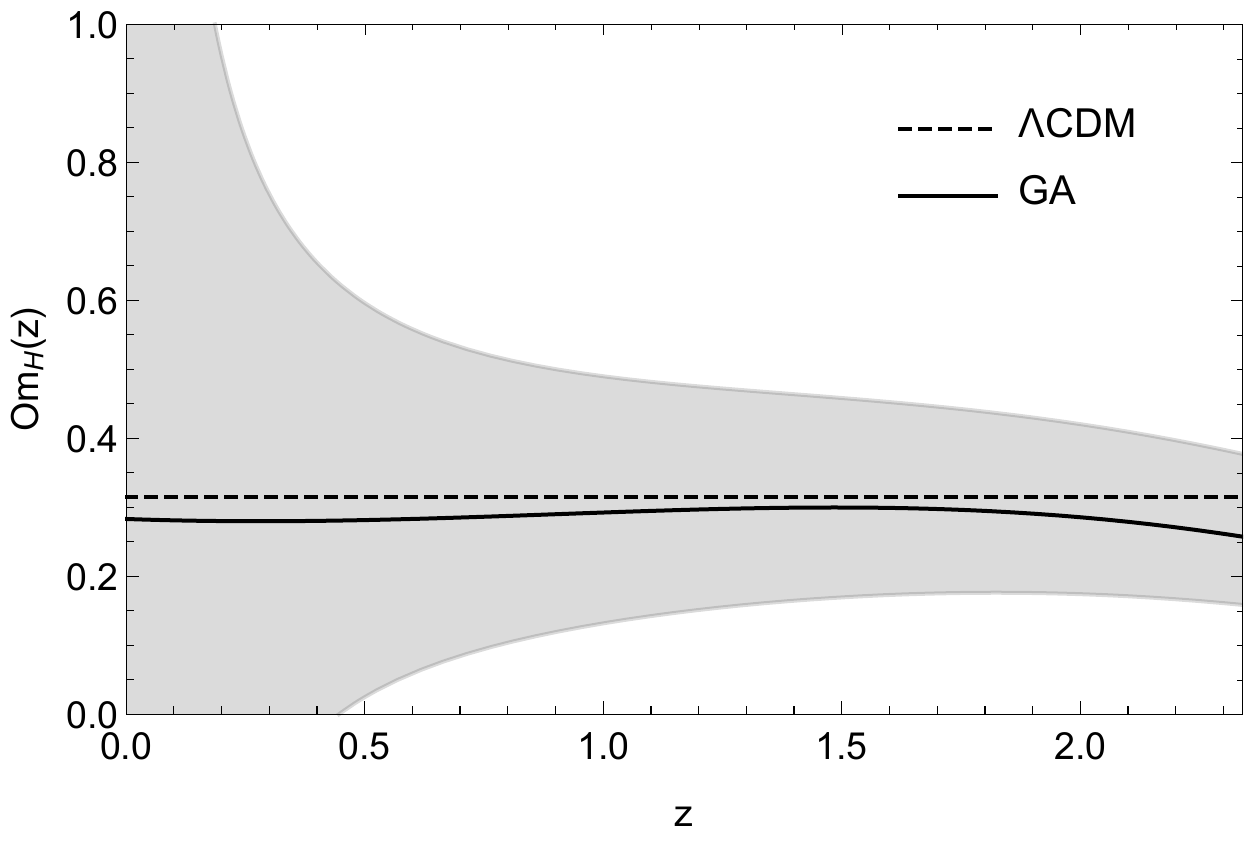}
\includegraphics[width = 0.49\textwidth]{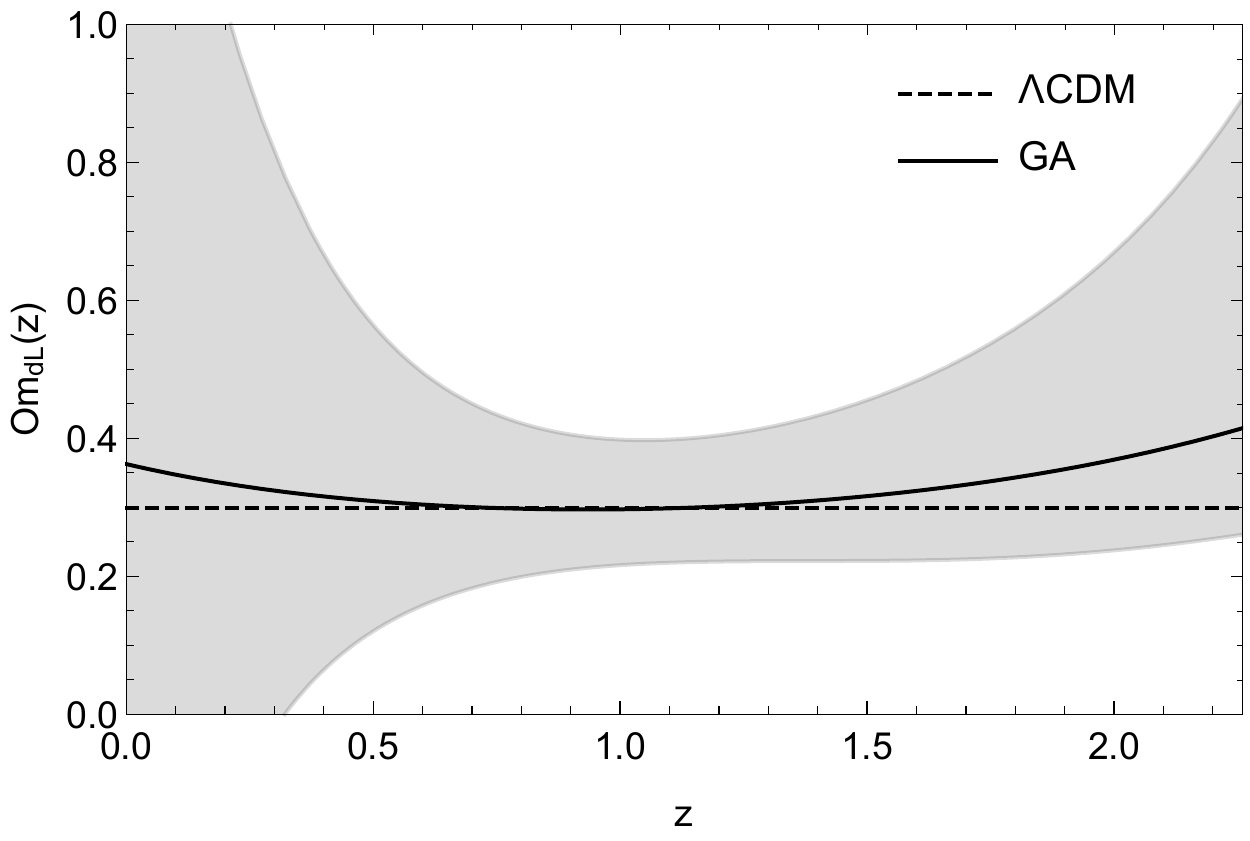}
\caption{The reconstruction of the $\textrm{Om}_\textrm{H}(z)$ (left) and $\textrm{Om}_\textrm{dL}(z)$ (right) statistics respectively, along with the $1\sigma$ errors (gray regions). Both cases are consistent with the \lcdm model (dashed line).\label{fig:Omstats}}
\end{figure*}

\subsection{The null tests \label{sec:null}}
Next we focus on the reconstruction of the null tests for the \lcdm model. The first null test we will consider is the $\textrm{Om}(z)$ statistic of Ref.~\cite{Sahni:2008xx}, which only requires knowledge of the Hubble parameter $H(z)$ and allows us to discriminate \lcdm from other DE models, see Refs~\cite{Nesseris:2010ep}. It is defined as\footnote{We use the notation $\textrm{Om}_\textrm{H}$ with the subscript $\textrm{H}$ to discriminate this null test from the one we will introduce later on and which is based on the luminosity distance $d_L(z)$.}
\be
\textrm{Om}_\textrm{H}(z)\equiv \frac{H(z)^2/H_0^2-1}{(1+z)^3-1}.\label{eq:om_statistic}
\ee

Here we also present a different, but at the same time complementary, null test of the \lcdm by extracting the matter density $\Omega_{m0}$ from the luminosity distance instead of the Hubble parameter. To do this, we use the Lagrange inversion theorem which states that given an analytic function, we can estimate the Taylor series expansion of the inverse function, i.e. given the function $y=f(x)$, where $f$ is analytic at a point $p$ and $f'(p)\neq 0$ the theorem allows us to solve the equation for $x$ and write it as a power series $x=g(y)$, see \cite{Abramowitz}.

We now apply the Lagrange inversion theorem to the luminosity distance $d_L(z,\Omega_{m0})$ and from now on we will restrict ourselves at late times, when DE dominates over the other components, such as radiation and neutrinos. Then, the analytical expression of the luminosity distance for the \lcdm model, assuming a flat Universe but neglecting radiation and neutrinos, is given by
{\small
\ba
d_L(z,\Omega_{m0})&=&\frac{c}{H_0}(1+z) \int_0^z\frac{1}{H(x)} dx \nn\\
&=&\frac{c}{H_0}\frac{2(1+z)}{\sqrt{\Omega_{m0}}}\left(_2F_1\left(\frac{1}{6},\frac{1}{2},\frac{7}{6},\frac{\Omega_{m0}-1}{\Omega_{m0}}\right)-\right.\nn\\
& &  \left. \frac{_2F_1\left(\frac{1}{6},\frac{1}{2},\frac{7}{6},\frac{\Omega_{m0}-1}{\Omega_{m0}(1+z)^{3}}\right)}{\sqrt{1+z}}\right).\label{eq:dl}
\ea}
To derive the $\textrm{Om}_\textrm{dL}(z)$ test we first do a series expansion on Eq.~\eqref{eq:dl} around $\Omega_{m0}=1$ and keep the first $10$ terms in order to obtain a reliable unbiased estimation and avoid theoretical systematic errors. Then, we apply the Lagrange inversion theorem to invert the series and to write the matter density $\Omega_{m0}$ as a function of the luminosity distance $d_L$, i.e $\textrm{Om}_\textrm{dL}=\textrm{Om}_\textrm{dL}(z,d_L)$. For example, the first two terms of the  expansion are
\be
\textrm{Om}_\textrm{dL}(a,d_L)=1-\frac{7a\left(\frac{H_0}{c}d_L-\frac{2-2\sqrt{a}}{a}\right)}{6+\sqrt{a}\left(a^3-7\right)}+\cdots,
\ee
where the scale factor $a$ is related to the redshift $z$ as $a=\frac{1}{1+z}$. This null test has the main advantage that it does not require taking derivatives of the data as we use the luminosity distance directly.

The reconstruction of both null tests of the \lcdm model is shown in Fig.~\ref{fig:Omstats}, in the left panel for the $\textrm{Om}_\textrm{H}$ and the right panel for the $\textrm{Om}_\textrm{dL}$ respectively. We find that both null tests are in agreement with  \lcdm at the $1\sigma$ level. While the errors of the distance modulus $\mu(z)$ and the $\textrm{Om}_\textrm{dL}$ test, shown in Figs.~
\ref{fig:Data} and \ref{fig:Omstats} respectively, seem somewhat larger compared to those in Refs.~\cite{Nesseris:2012tt,Nesseris:2010ep}, the latter used the Union 2.1 set but did not include the systematic errors, thus underestimating the errors regions. On the other hand, the Pantheon compilation both statistical and systematic errors are included in the publicly available data \footnote{The systematic errors are included in the ``sys\_full\_long.txt" file, which is publicly available from the Pantheon GitHub page \href{https://github.com/dscolnic/Pantheon}{https://github.com/dscolnic/Pantheon}. For the SnIa likelihood in our analysis, we use the free and publicly available code written by one of the authors, see \href{https://members.ift.uam-csic.es/savvas.nesseris/codes.html}{https://members.ift.uam-csic.es/savvas.nesseris/codes.html}.}.
As a result, even though the Pantheon set has roughly twice as many points than the Union 2.1, the inclusion of the systematic errors of the Pantheon in the analysis, brings the error estimates for $\mu(z)$ and $\textrm{Om}_\textrm{dL}$ to the same level as those in Refs.~\cite{Nesseris:2012tt,Nesseris:2010ep}.

\section{Conclusions \label{sec:conclusions}}
In summary, ML methods are revolutionizing the way we interpret data since they can help to remove biases due to choosing \emph{a priori} a specific defined model. This is more important than ever as the endeavor to explain the accelerated expansion of the Universe has led to a plethora of DE models, which make the interpretation of the data difficult as the results are model dependent. This can lead to model bias, thus affecting the conclusions drawn about fundamental physics.

We have shown that applying the GA to the SnIa and $H(z)$ data can be used to reconstruct the expansion history of the Universe and help determine the current deceleration parameter and transition redshift in a model independent fashion. The datasets we use are the Pantheon Type Ia Supernovae compilation of Ref.~\cite{Scolnic:2017caz} and the $H(z)$ based on the differential age method and the clustering of galaxies or quasars by Moresco \emph{et al.} (shown in Table I of Ref.~\cite{Arjona:2018jhh}), both being state-of-the-art at the moment. Given that we only have one realization of ``real" data at the moment, one could possibly use mock datasets to test the GA approach as a reconstruction method. This however, has already been done, see for example \cite{Nesseris:2012tt,Nesseris:2014qca}. By considering subsamples of the Pantheon dataset at high redshifts we also confirmed that our results are robust.

We also find a $\sim4.5\sigma$ detection of the accelerated expansion, contrary to recent claims by Ref.~\cite{Nielsen:2015pga, Colin:2018ghy}, where the authors claimed that there is little to no evidence for acceleration. The main differences between our work and that of Ref.~\cite{Nielsen:2015pga}, is that in the latter the authors used the (now outdated) Joint Lightcurve Analysis (JLA) catalogue by Ref.~\cite{Betoule:2014frx}, while here we use the much more recent Pantheon sample by Ref.~\cite{Scolnic:2017caz}. The Pantheon sample was created by analyzing together recent observations of SnIa from the Pan-STARRS1 survey and from other previously available low redshift subsamples from other surveys, in order to create a uniform dataset that would have the same quality cuts and systematics.

Besides the choice of the SnIa data, our paper and that of Ref.~\cite{Nielsen:2015pga}, also differ in the fact that while our approach is completely nonparametric and model-independent, Ref.~\cite{Nielsen:2015pga} assumes Gaussian priors for the absolute magnitude $M$, the stretch $x$ and the color $c$, as seen in Eq.~(4) in their paper, each with a mean value and a standard deviation. Then, these six new parameters are fitted along with the cosmological parameters. However, as was pointed out in \cite{Rubin:2016iqe}, the observed distributions of these parameters are far from redshift-independent, thus biasing their results.

Furthermore, our method has several advantages compared to other methods found in the literature like the GP. In particular, while the GP requires the choice of a kernel function and a fiducial model, usually taken to be a Gaussian and \lcdm respectively, our approach assumes neither and is completely theory agnostic. Also, compared to the approach of Ref.~\cite{Nielsen:2015pga}, our approach is completely nonparametric.

In summary, we showed how the GA can be used to reconstruct complementary null tests of the \lcdm model via reconstructions of both the Hubble parameter and the luminosity distance and we found that both are consistent with \lcdm within the errors.

\section*{Acknowledgements}
It is our pleasure to thank Juan Garc\'{\i}a-Bellido and Arman Shafieloo for enlightening discussions. The authors acknowledge support from the research project No. PGC2018-094773-B-C32 and the Centro de Excelencia Severo Ochoa Program SEV-2016-0597. S.~N. also acknowledges support from the Ram\'{o}n y Cajal program through Grant No. RYC-2014-15843. \\

\textbf{Numerical Analysis Files}: The numerical codes used by the authors in this paper will be released upon publication at \cite{mywebsite}, but also at the GitHub repositories  \cite{SNgit} and \cite{RAgit}.

\bibliography{null_test}

\end{document}